\documentclass[12pt]{article}
\usepackage{blois,graphicx}

\begin{document}
\title{The Gaps in Double Diffraction Events at LHC and Baryon-Antibaryon Torus}
\author{Olga Piskounova}
\address{P.N.Lebedev Physics Institute of Russian Academy of Science, Leninski prosp. 53, 119991 Moscow, Russia}

\maketitle\abstract{
The positive baryon-antibaryon production asymmetries that have been measured at LHC are real demonstrations of string junction dynamics in the proton-proton interactions. 
The topological presentation of pomeron exchange in the proton-proton collision of high energy is cylinder that is covered with quark-gluon net. I assume that the process of double diffraction (DD) can be presented as a diagram of one pomeron exchange with the central loop of two pomeron cylinders, that is similar to the pomeron diagram with the handle or the pomeron torus. Taking into account that the junction of three gluons (SJ) has the positive baryon number, as well as the antijunction is of negative baryon charge, our pomeron construction can be covered by only a certain number of hexagons with 3 string junction and 3 antijunction vertices each.  It is reasonable to expect that the dynamics of rapidity gaps in DD should be determined by the number of hexagons on the surface of pomeron torus. Therefore, the gap distribution in DD events has some discrete structure in the region of large gaps. Moreover, the string-junction conglomerate could be released after DD interaction as a metastable particle. There is another multiparticle process 
with pomeron loop configuration that is event with doubled multiplicity in the central rapidity gap, which corresponds to the cut along the pomeron loop. The events with the gap in DD as well as the events with doubled multiplicity, each go on the level of 1.2% of the inclusive production cross section.}

%\def\be{\begin{equation}}
%\def\ee{\end{equation}}
%\def\bea{\begin{eqnarray}}
%\def\eea{\end{eqnarray}}
%\def\met{\not{\!{\rm E}}_T}
%\def\zp{Z^\prime}

%\newcommand{\bi}{\bibitem}
%\newcommand{\nn}{\nonumber}
%\def\BibTeX{\rm B{\sc ib}\TeX}

%\begin{document}

%\title{The Gaps in Double Diffraction Events at LHC as Manifestation of Pomeron Torus Exchange}

%\author{Olga Piskounova,\auno{1}}
%\address{$^1$P.N.Lebedev Physics Institute of Russian Academy of Science, Leninski prosp. 53, 119991 Moscow, Russia}
 
%\begin{abstract}
%\end{abstract}

%\maketitle

%\begin{keyword}
Keywords: Multiparticle production, Pomeron exchange, LHC, Double Diffraction, Pomeron Torus diagram, String Junction}
%\end{keyword}

\section{Introduction}
\subsection*{Measurements of Rapidity Gaps at the Double Diffraction Dissociation}
\begin{figure}[htpb]
\centering
  \includegraphics[scale=0.5]{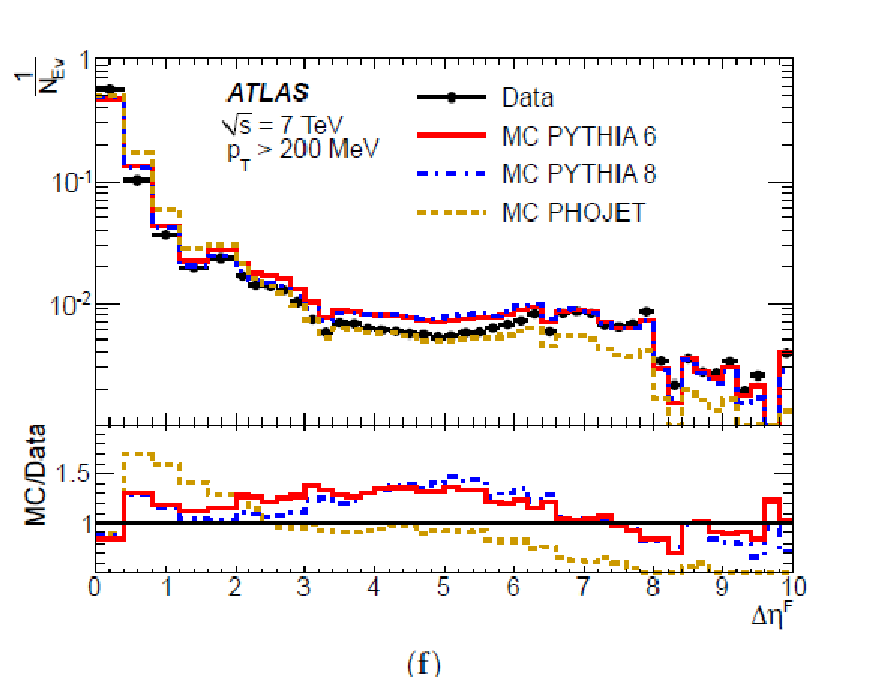}
  \caption{Forward rapidity gap distribution in ATLAS experiment.}
  \label{atlas}
\end{figure}

The measurement of diffraction gaps in ATLAS experiment figure~\ref{atlas} at LHC \cite{atlas} has shown that the behavior of their distribution has different character in the different gap ranges. The histogram at the large values of gap indicates some discrete states of gaps. In this area only the process of double diffraction dissociation (DD) gives the contribution into the histogram. Of course, the discrete pattern can be initiated by poor statistics. Nevertheless, we have to study here few diagrams that may lead to discrete levels of DD gap.

\subsection*{Topological Expansion and Pomeron Exchange}
\begin{figure}[htpb]
\centering
  \includegraphics[scale=0.5]{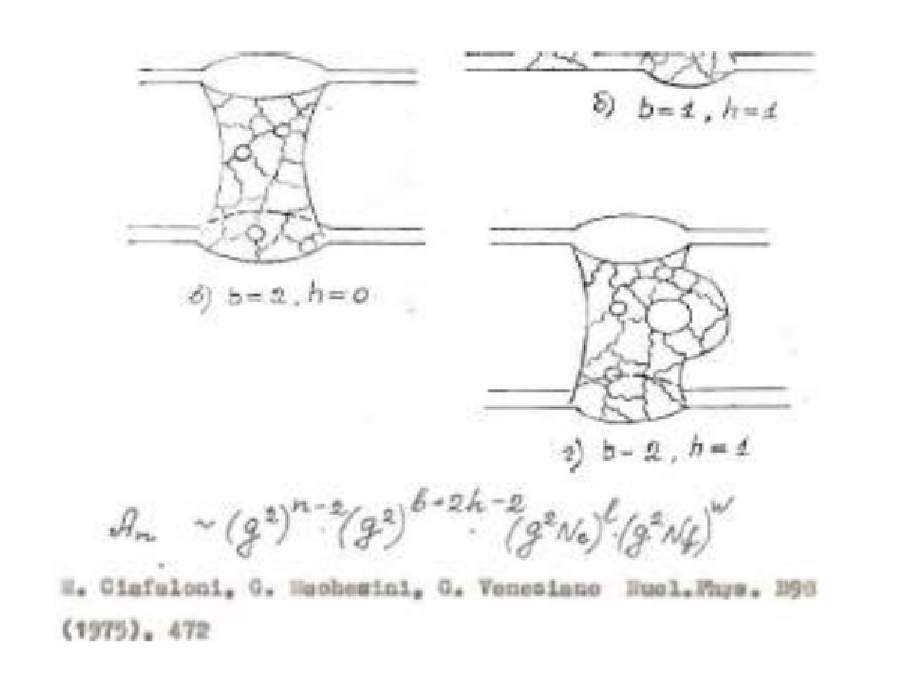}
  \caption{The fragment of graphical presentation of topological expansion, where b is the number of
boundaries and h is the number of handles.}
  \label{myPhD}
\end{figure}

Pomeron exchange is the topological QCD diagram that is responsible for multi particle production in p-p collisions at LHC energies. It was drawn in the topological expansion \cite{topologyexp} as the cylindrical net of gluon exchanges with the random amount of quark-antiquark loops inserted between them. The topological expansion gives the chance to classify the contributions from general diagrams of multi particle production in the hadron interactions. This expansion has practically allowed us to build Quark-Gluon String Model (QGSM) \cite{difreport,QGSM,nuclphys,recentpub}. Few orders in topological expansion are graphically presented in the figure from my PhD thesis figure~\ref {myPhD}, where the third order is named pomeron with the handle. Double diffraction dissociation in this presentation looks like the cylinder of one pomeron exchange with the cylindrical handle \cite{topologyexp, myPhD}.  Pomeron exchange used to take 1/9 from the leading contribution of planar diagram with quark-antiquark annihilation on the energy scale $\sqrt{s}$ is of the order 1 GeV. But the quark annihilation diagram is dying out with energy as $s^{-0.5}$ due to the regge behavior of quark-gluon diagrams. In the same time the cross sections of one pomeron exchange is to be growing as $s^{\Delta_P}$, where the pomeron trajectory parameter $\Delta_P = \alpha_P(0)-1$ = 0.12. Such a way, the second order diagrams with the pomeron exchange have to be dominating at high energies. 

\subsection{String Junction and the Positive Baryon Production Asymmetry at LHC}
Three gluon connection of Mercedes type, or String Junctions (SJ), plays an important role in the multi particle production in our positive-baryon-charge world as well as in p-p collisions of high energy. This object brings the positive baryon charge and generates the great asymmetry between baryon and antibaryon spectra in the region of diquark fragmentation. It is responsible also for the baryon production asymmetry at the central rapidity point in any collisions with proton matter. This asymmetry is surviving even at LHC energies that is a clear evidence of SJ transfer upto the central rapidities that has been confirmed by available collider data \cite{SJ,baryonasymmetry}. It will be shown in this article that the dynamics of DD gaps is obviously ruled by the regge parameters of SJ trajectory too.
 
\section{ Double Diffractive Dissociation as an Exchange with Pomeron Torus}
\begin{figure}[htpb]
\centering
  \includegraphics[scale=0.5 ]{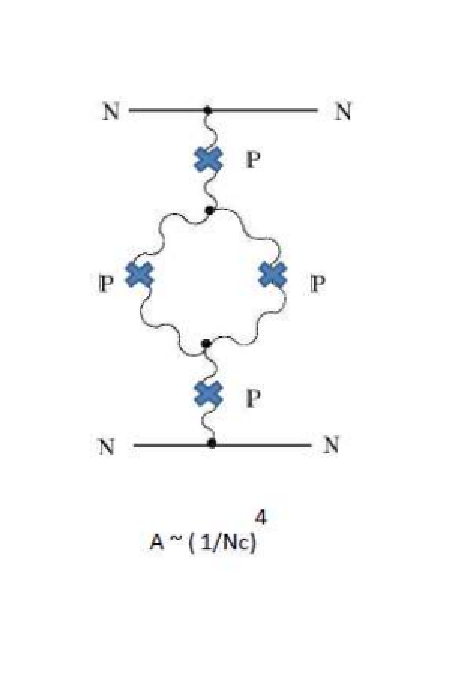}
  \caption{Two-pomeron loop in the center of one-pomeron exchange.}
  \label{DD}
\end{figure}

Double diffraction dissociation (DD) is a next order in the topological expansion after the pomeron exchange and should be presented as one pomeron exchange with pomeron loop in the center, see figure~\ref{DD}.
 
Actually, the topological cylinder with the handle takes $(1/9)^2$ percent from the pomeron exchange cross section, which is approximately equal to 1,2% of inelastic cross section at LHC energies. If the central pomeron loop was not cut, we have spesific spectra of produced hadrons: two intervals at the ends of rapidity range that are populated with multi particle production, and the valuable gap in the center of rapidity, see figure~\ref{multiplicity_with_gap}.

\begin{figure}[htpb]
\centering
  \includegraphics[scale=0.5]{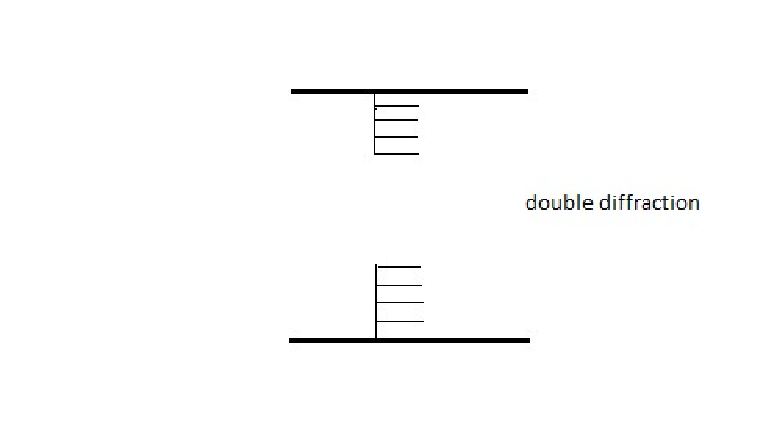}
  \caption{The diagram of particle multiplicity in the events with central gap.}
  \label{multiplicity_with_gap}
\end{figure}

Looking at the two-pomeron loop in the diagram, we are realizing that it is torus in 3D topology. This interesting object should be considered separately in order to reveal some remarkable features for the experimental detection.
\section{Junction-Antijunction Hexagon Net and Discrete Dynamics of DD Gaps}
\begin{figure}[htpb]
\centering
  \includegraphics[scale=0.37]{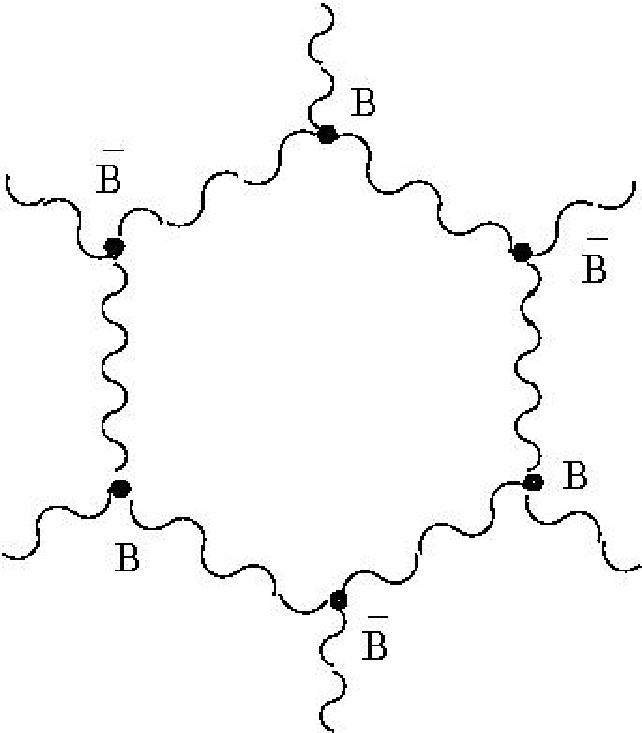}
  \caption{One cell of hexagon net with the string and anti-string junctions.}
  \label{onecell}
\end{figure}

As we remember the pomeron cylinder is built by gluon exchange net, let us consider only three gluon connections on the surface of torus (or pomeron cylinder loop). This String-Junction type of gluon vertices has been studied in our early researches \cite {kaidalov,baryonasymmetry}. Since this object brings the baryon charge, the antiSJ also exists and brings the charge of antibaryon. The only charge-neutral way to construct the net from SJs and antiSJs is hexagon where antibaryon charge is following the baryon one as it is shown in the figure~\ref{onecell}. 
 
\begin{figure}[htpb]
\centering
  \includegraphics[scale=0.24]{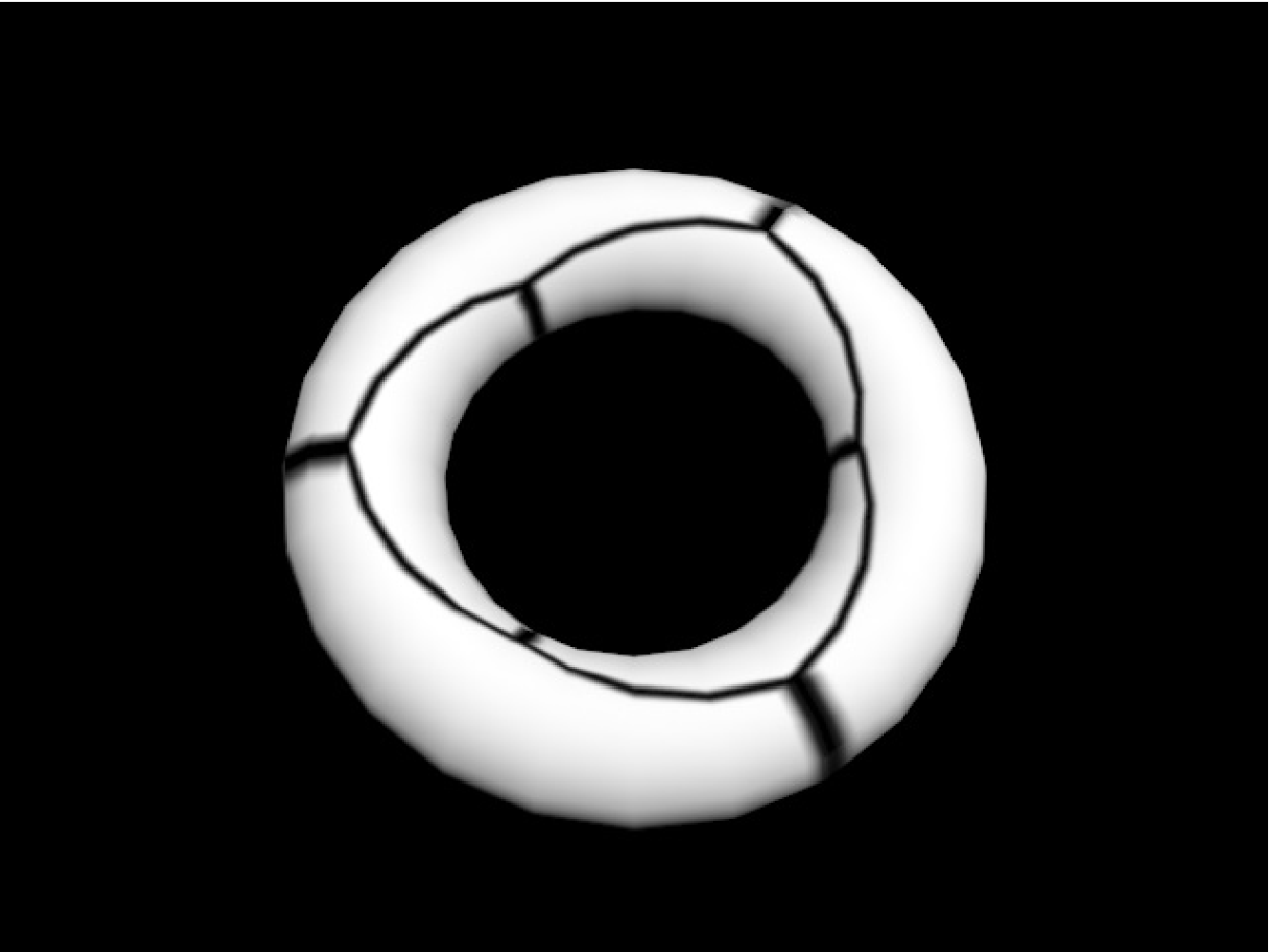}
  \caption{Six hexagons on the surface of pomeron torus.}
  \label{torus}
\end{figure}

The closed net of six hexagons on the torus is shown in the figure~\ref{torus}.

\begin{figure}[htpb]
\centering
  \includegraphics[scale=0.5]{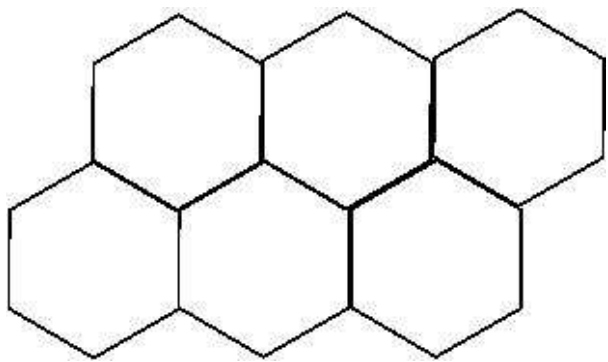}
  \caption{An example of six-hexagon net, which can be closed on the torus.}
  \label{sixbee}
\end{figure}

If people are trying to match the eligible number of hexagons, it becomes clear that there is a discrete row: Hexnumb = 4, 6, 8, 12, 16, 24, 32, 48 etc, see figure~\ref{sixbee}. It means that the pomeron torus has certain levels of energy. Such a discourse leads to discrete gap states at DD.
 
\section{Doubled Multiplicity at the Inclusive Hadroproduction as a Result of Pomeron Torus Exchange}
We have to learn here also the pattern with the torus that is cut through both sides, because this case of multi particle production is a version of the same diagram as in figure~/ref{DD}. The diagram figure~\ref{doublemult} shows clearly that the same gap, which takes place in DD, is filled with particles of doubled density due to both sides of cut torus that give twice as more particles per rapidity unit than one pomeron exchange produces. It is another evidence of our pomeron loop in multi particle production statistics. The events of the gap filled with doubled multiplicity should have the same distributions as the empty gaps in DD statistics. 

\begin{figure}[htpb]
\centering
  \includegraphics[scale=0.5]{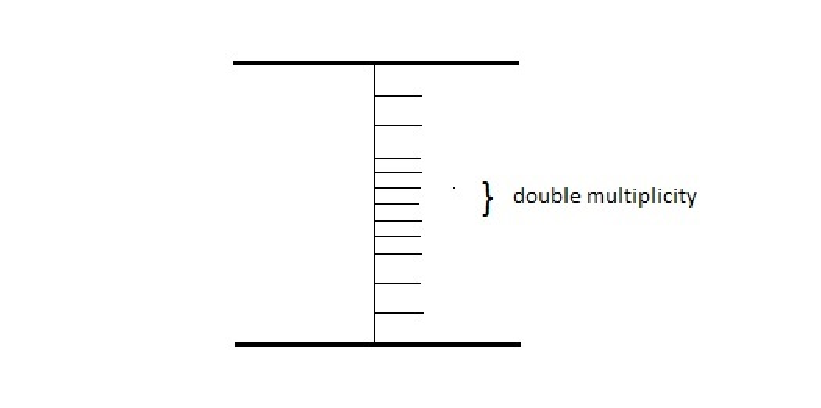}
  \caption{The particle production diagram with doubled density in the central rapidity gap.}
  \label{doublemult.}
\end{figure}

\section{More Suggestions on the Pomeron Torus}
Here it is the place to imagine where to the pomeron torus could contribute. What we have, if our gluon-junction construction that looks like a compactificated pomeron string would be released as metastable particle? It is charge neutral QCD cluster with the certain potential energy, which is determined by number of hexagons, Hexnumb = (NSL + NantiSJ)/2. If such cluster would be stable in an environment with high pressure of gravitation, this is appropriate candidate for the dark matter (DM)\cite{torus,aspen}.  The reasons, why this object does not dissipate in the collisions with matter, are the following: pomeron torus have valuable energy of connection and, if the atomic number of elements in the space is too small in comparison with the number of SJ-antiSJ vertices in our toroidal construction, this compact torus leaves intact after the collisions with the less heavy atoms. Because of high energy  collision with protons and nuclei in the space or wherever it takes place, some amount of energy of pomeron torus could also dissipate into baryons and mesons. 

\section{Conclusion}
The topological presentation of pomeron exchange at the proton-proton collision of high energy is cylinder that is covered with quark-gluon net. I suggest that the process of double diffraction (DD) can be presented as one pomeron exchange with the central loop of two uncut pomeron cylinders.  Taking into account that the junction of three gluons (SJ) has the positive baryon number, as well as the antijunction is of negative baryon charge, our pomeron construction can be covered by only a certain number of hexagons with 3 string junction and 3 antijunction vertices each.  It is reasonable to expect that the dynamics of rapidity gaps in DD should be determined by the number of hexagons on the surface of pomeron torus. Therefore, the gap distribution in DD events has the discrete structure in the region of large gaps. Moreover, the string-junction torus can be released in the course of pp interaction as metastable particle. The possibility of production of the states with many string junctions has been discussed recently by G.C. Rossi and G. Veneziano \cite{newveneziano}. There is another process with pomeron loop configuration that is the particle production events with doubled multiplicity at the central rapidity, which corresponds to the cut along the pomeron loop. The positive hyperon production asymmetries that have been measured at LHC are more practical demonstration of string junction dynamics in proton-proton interactions. These measurements have shown that the energy dependence of baryon excess in the central rapidity region depends on the parameter of the string junction. Therefore, the actual goal of LHC experiments should be the measurement of baryon-antibaryon asymmetry in heavy baryon production that helps us to estimate the important dynamical parameter of string junction, the intercept of SJ-antiSJ Regge trajectory. Furthermore, the detail study of the gap distributions in DD as well as the characteristics of events with doubled multiplicity, which both go on the level of 1.2% of total production cross section, seems also advise.

\section*{Acknowledgements}
Author would like to express her gratitude to Prof. O.Kancheli for numerous discussions and to Vladimir Tskhay for the designing of the figure with torus. The part of this work has been performed at Aspen winter school \cite{aspen}, which is supported by National Science Foundation grant PHY-1607611.

%\newpage
\bibliographystyle{unsrt}

\begin{thebibliography}{99}
\bibitem{atlas}ATLAS collab., Eur.Phys.J. {\bf C72}, 1926 (2013), arXiv:1201.2808.
\bibitem{topologyexp} M Ciafaloni, G. Marchesini, G. Veneziano, Nucl. Phys. {\bf B98}, 472 (1975).
\bibitem{difreport} A.Kaidalov, Phys. Rep. {\bf 50}, 1 (1979).
\bibitem{QGSM} A. Kaidalov and K. Ter-Martirosyan, peprint ITEP {\bf 161}, 1 (1983).
\bibitem{nuclphys} A.Kaidalov and O. Piskunova, Sov. Jou. Nucl Phys. {\bf 43}, 1545 (1985).
\bibitem{kaidalov} A. Kaidalov and O. Piskunova , Z.Phys. {\bf C30}, 145 (1986).
\bibitem{myPhD} O. Piskunova, PhD thesis (1988), in Russian.
\bibitem{SJ} O. Piskunova, Phys. Atom. Nucl. {\bf 70} 1107 (2007), hep-ph/0604157 (2006). 
\bibitem{baryonasymmetry} M.A. Erofeeva and O.I. Piskounova,  Nonlin.Phenom.Complex Syst. {\bf 12}, 425 (2009).  
\bibitem{myatlastalk} O. Piskounova, the talk at ATLAS Diffraction Working Group meeting, CERN, 20 April 2015.
\bibitem{newveneziano} G. Rossi and G. Veneziano, JHEP {\bf 1606}, 41 (2016), arxiv:
\bibitem{ICHEP} O.Piskounova, poster presentation, PoS ICHEP2016, 711 (2017).
\bibitem{torus} O. Piskounova, my talk at LHC Forward Phys. and Diffraction, CERN, 2018, arxiv:1702.02769, arXiv:1812.02691.  
\bibitem{recentpub} O. Piskounova, Int. Jou. of Mod. Phys. {\bf A35}, 2050067 (2020) , arXiv:1706.07648. 
\bibitem{aspen} O. Piskounova, Poster presentation at 'A Rainbow of Dark Sectors' Conference, Aspen Center for Physics, 2th April, 2021.
\end{thebibliography}

\end{document}